%% file: net-emb-init.tex
\documentclass[sigconf]{acmart} 

\usepackage{balance}

\def\BibTeX{{\rm B\kern-.05em{\sc i\kern-.025em b}\kern-.08emT\kern-.1667em\lower.7ex\hbox{E}\kern-.125emX}}

\copyrightyear{2020} 
\acmYear{2020} 
\setcopyright{acmcopyright}
\acmConference[WSDM '20]{The Thirteenth ACM International Conference on Web Search and Data Mining}{February 3--7, 2020}{Houston, TX, USA}
\acmBooktitle{The Thirteenth ACM International Conference on Web Search and Data Mining (WSDM '20), February 3--7, 2020, Houston, TX, USA}
\acmPrice{15.00}
\acmDOI{10.1145/3336191.3371781}
\acmISBN{978-1-4503-6822-3/20/02}

\usepackage{multirow}
\usepackage[ruled, vlined, linesnumbered]{algorithm2e}

\input{macro}

\begin{document}

\fancyhead{}

\title{Initialization for Network Embedding: A Graph Partition Approach}

\author{Wenqing Lin}
\email{edwlin@tencent.com}
\affiliation{%
  \institution{Tencent Inc.}
  \state{Shenzhen, China}
}

\author{Feng He}
\email{fenghe@tencent.com}
\affiliation{%
  \institution{Tencent Inc.}
  \state{Shenzhen, China}
}

\author{Faqiang Zhang}
\email{fankyzhang@tencent.com}
\affiliation{%
  \institution{Tencent Inc.}
  \state{Shenzhen, China}
}

\author{Xu Cheng}
\email{alexcheng@tencent.com}
\affiliation{%
  \institution{Tencent Inc.}
  \state{Shenzhen, China}
}

\author{Hongyun Cai}
\email{laineycai@tencent.com}
\affiliation{%
  \institution{Tencent Inc.}
  \state{Shenzhen, China}
}

\begin{abstract}
 Network embedding has been intensively studied in the literature and widely used in various applications, such as link prediction and node classification. While previous work focus on the design of new algorithms or are tailored for various problem settings, the discussion of initialization strategies in the learning process is often missed. In this work, we address this important issue of initialization for network embedding that could dramatically improve the performance of the algorithms on both effectiveness and efficiency. Specifically, we first exploit the graph partition technique that divides the graph into several disjoint subsets, and then construct an abstract graph based on the partitions. We obtain the initialization of the embedding for each node in the graph by computing the network embedding on the abstract graph, which is much smaller than the input graph, and then propagating the embedding among the nodes in the input graph. With extensive experiments on various datasets, we demonstrate that our initialization technique significantly improves the performance of the state-of-the-art algorithms on the evaluations of link prediction and node classification by up to 7.76\% and 8.74\% respectively. Besides, we show that the technique of initialization reduces the running time of the state-of-the-arts by at least 20\%.
\end{abstract}


%
%


%

\renewcommand{\textrightarrow}{$\rightarrow$}

\maketitle

\input{introduction}

\input{preliminaries}

\input{overview}

\input{solution}

\input{experiments}

\input{related}

\input{conclusions}

\begin{balance}
\bibliographystyle{ACM-Reference-Format}
\bibliography{ref}
\end{balance}

\end{document}

%% file: macro.tex
\def\header{\vspace{1.5mm} \noindent}

\def\figcapup{\vspace{-2mm}}

\def\done{\vspace{-0.5mm} \hspace*{\fill} {$\square$}}

\def\nbr{N}

\newtheorem{example}{Example}

%% file: introduction.tex
\section{Introduction} \label{sec:introduction}

Graphs are so ubiquitous that most of data can be naturally modeled as graphs, not to mention the social networks. Network embedding \cite{hvk17,pxjw17,wrj17,djxc18,pe18,hbrs18} is an intensively studied and widely used technique, which assigns each node in the graph a fixed-length vector that preserves the structure of graph and is helpful in various tasks, such as link prediction and node classification. As such, network embedding alleviates the difficult issue of feature engineering on the graph. The solutions to network embedding can be roughly classified into two categories, namely random walk based approaches \cite{brs14,aj16} and matrix based approaches \cite{swq15,dpw16}. 

However, the problem of network embedding is non-convex \cite{hbys18} rendering the previous approaches rely on the stochastic gradient descent (SGD) technique for optimization, which would incur the issue of stuckness in the local minima. Therefore, the initialization strategies in the learning of network embedding, that takes into account the structure of the input graph, would dramatically affect the performance of the network embedding algorithms.

\begin{figure*}[t]
\centering
\begin{tabular}{cc}
 \hspace{-2mm} \includegraphics[height = 25mm]{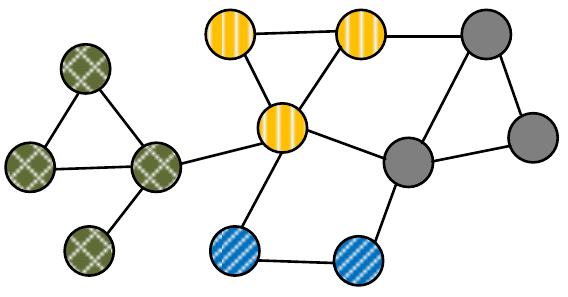} &
 \hspace{4mm} \includegraphics[height = 25mm]{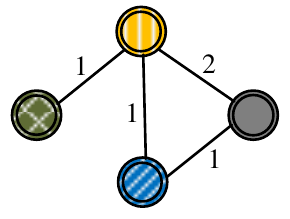} \\
 \hspace{-2mm} (a) $G$. & \hspace{4mm} (b) $G_a$.  \\
\end{tabular}
\caption{A graph $G$ with $4$ partitions, each of which is colored differently, and the abstract graph $G_a$ of $G$.}  \label{fig:abstract-graph}
\end{figure*}

The previous approaches \cite{hbys18} for the initialization in the computation of network embedding take two steps: First, they coarsen the edges or the star structures of the input graph $G$ which produces a smaller graph $g$; Then, they exploit the existing algorithms \cite{brs14,aj16,swq15,dpw16} to compute $g$'s network embedding, which are directly used as the initialization in the learning of $G$'s network embedding. However, there exist some issues that would make these approaches deficient. Firstly, the coarsening method considers only the local structure, which might not reflect the overall structure of the input graph. For example, an edge playing the role of bridge \cite{r74} in the graph could be coarsened, rendering the communities incident to the bridge even difficult to be separated from each other. Secondly, since a node $v$ in $G$ might be pertinent to multiple nodes in $g$, the direct inheritance of the embedding from one node in $g$ would result in the missing of $v$'s important structural features in $G$. Thirdly, there exist several hyperparameters in the existing algorithms \cite{brs14,aj16,swq15,dpw16}, which would highly degrade their performance without careful configuration. However, the previous approaches do not provide any effective solution about the tuning of hyperparameters.

To address the aforementioned issues in the previous approaches, we propose a graph partition based algorithm, dubbed as GPA, which first divides the input graph $G$ into several disjoint subsets by the graph partition algorithm \cite{gv98} that minimizes the edge cut between subsets. Based on that, we collapse the subgraph induced on each partition as an abstract node and the cutting edges as the weighted edges to construct an abstract graph $G_a$, which is of size much smaller than $G$ and represents the sketch of $G$. Afterwards, we compute the network embedding of $G_a$ by a modified version of the existing network embedding algorithm \cite{brs14}. 
Note that, it is highly costly to tune the hyperparameters of the network embedding algorithm on the fly, due to the huge search space and expensive evaluation cost. To alleviate this issue, we devise an approach that learns a regression model for the hyperparameter configurations in a preprocessing step and computes a suitable configuration in linear time. Finally, the initial embedding of each node in $G$ is computed by propagating $G_a$'s embedding among the nodes in $G$. In the experiments, we demonstrate that the performance of GPA outperforms the state-of-the-arts on various tasks, i.e., link prediction and node classification. Besides, we show that the initialization strategies of GPA lead to the speedup of the running time of the baseline algorithms.

In summary, the contributions of the present work are the followings.
\begin{itemize}
  \item We devise the GPA algorithm as an effective technique for the initialization of network embedding algorithms. Specifically, GPA considers the structure of the input graph by exploiting the graph partition algorithm to construct the sketch of a graph and minimize the size of edge cut.
  \item We develop the algorithm to generate the abstract graph, which is a weighted graph and is much smaller than the input graph. We also devise the algorithm to compute the network embedding on the weighted graph, which is not discussed in the previous approaches.
  \item We propose an efficient algorithm that produces the initial embedding of each node in the input graph from the embedding of the abstract graph, and smooths the initialization via a propagation process.
  \item We develop the hyperparameter learning algorithm that addresses the issue of hyperparameter tuning for the network embedding on the abstract graph, which improves the performance of the proposed algorithm.
  \item We demonstrate in various experiments where GPA outperforms the state-of-the-arts by up to 8.74\% performance gain on effectiveness and reduces the running time by at least 20\%.
\end{itemize}

\header
{\bf Paper organization}. Section~\ref{sec:preliminaries} explains the definitions and notations used in the paper. Section~\ref{sec:overview} provides an overview of our solution, as well as the details of the algorithms that address the goal in this paper. After that, we demonstrate the superior performance of our algorithms compared with the baseline methods over several graphs. Finally, we discuss the related work in Section~\ref{sec:related} and conclude the paper in Section~\ref{sec:conclusions}.

%% file: preliminaries.tex
\section{Preliminaries} \label{sec:preliminaries}

Consider a graph $G=(V,E)$, where $V$ is the set of nodes and $E$ is the set of edges. We say that a node $v \in V$ is a {\it neighbor} of the other node $u \in V$ if there exists an edge $(u, v) \in E$. We denote $\nbr(v)$ as the set of neighbors of $v$ in $V$, i.e., $\nbr(v) \subseteq V$.

A {\it partitioning} of $G$, denoted by $\mathcal{P} = \{ V_1, V_2, \cdots, V_k \}$, divides $V$ into $k$ disjoint subsets where $k$ is a user-defined number, such that we have (i) $V_i \cap V_j = \emptyset$ where $1 \leq i < j \leq k$, and (ii) $\cup_{V' \in \mathcal{P}} V' = V$.

An {\it abstract graph} $G_a = (V_a, E_a)$ of $G$ is constructed on the partitioning $\mathcal{P}$ of $G$. In particular, each subset in $\mathcal{P}$ is represented as an {\it abstract node} $u_a$ in $V_a$. In other words, there is a {\it bijective function} $b$ that maps each partition $V' \in \mathcal{P}$ to an abstract node $u_a \in V_a$, i.e., $b(V') = u_a$. Besides, there is a {\it surjective function} $p$ that maps each node $v \in V$ to an abstract node $u_a \in V_a$, denoted by $p(v) = u_a$. In addition, we construct a {\it weighted} edge $(u_a, u_a') \in E_a$ for any two abstract nodes $u_a$ and $u_a'$ in $V_a$ if and only if there exist two nodes $v$ and $v'$ in $V$ such that we have (i) $p(v) = u_a$, (ii) $p(v') = u_a'$, and (iii) $(v, v') \in E$. The weight of $(u_a, u_a')$, denoted by $w(u_a, u_a')$, is computed as the number of such edges $(v, v')$. That is, a weighted edge in $G_a$ represents the edges in $G$ that connect the corresponding partitions.

\begin{example}
Figure~\ref{fig:abstract-graph}(a) shows a graph $G$ with $12$ nodes and $16$ edges. Assume that we partition $G$ into $4$ subsets, each of which is colored differently. Then, the nodes with the same color are collapsed as an abstract node. Therefore, there are $4$ abstract nodes in the abstract graph $G_a$ of $G$, as shown in Figure~\ref{fig:abstract-graph}(b). Besides, there is an edge of weight $2$ between the yellow abstract node and the gray abstract node in $G_a$, since there exist $2$ edges, each of which connects a yellow node and a gray node in $G$.
\done
\end{example}

\begin{figure*}[t]
\centering
\begin{tabular}{c}
\hspace{-5mm} \includegraphics[height = 33mm]{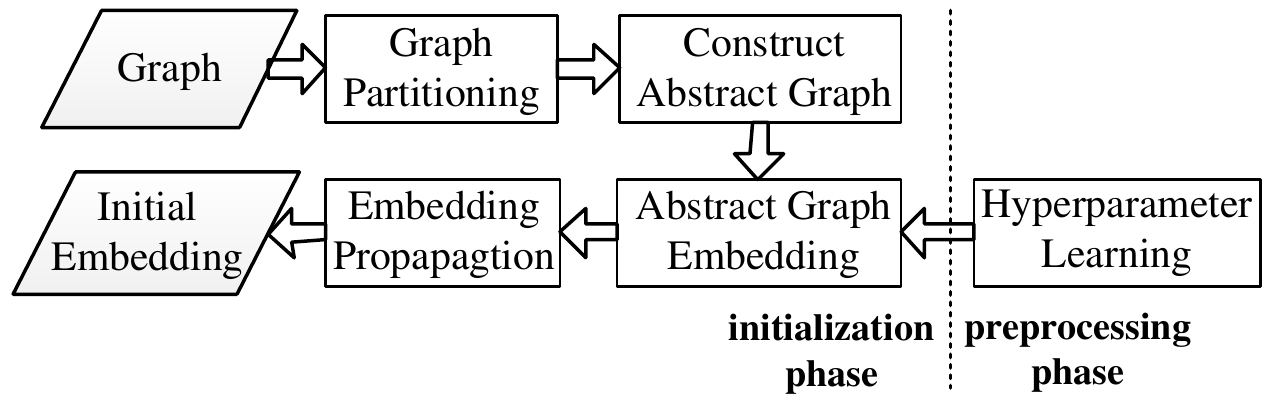}
\end{tabular}
\caption{The computing framework of GPA.} \label{fig:framework}
\figcapup
\end{figure*}

Given the graph $G=(V,E)$, the {\it network embedding} of $G$ maps each node $v \in V$ to a $d$-dimensional vector $f(v)$, where $f: V \rightarrow \mathbb{R}^d$ and $d$ is a user-defined parameter satisfying $d \ll |V|$. In general, network embedding should preserve the structure of $G$. In the other words, network embedding minimizes
\vspace{-2mm}
\begin{equation} \label{eqn:goal}
\sum_{v,u \in V} \big( \boldsymbol{A}_{v,u} - \theta (f(v), f(u)) \big)^2
\end{equation}
where $\boldsymbol{A} \in \mathbb{R}^{|V| \times |V|}$ could be the matrix of connections, such as the adjacency matrix of $G$, i.e., $\boldsymbol{A}_{v,u}$ is $1$ if $(v, u) \in E$ otherwise $0$, and $\theta$ is a similarity function that maps $f(v)$ and $f(u)$ to a real value in $\mathbb{R}$.

As aforementioned, most of the algorithms for network embedding ultimately exploit the technique of stochastic gradient descent (SGD) for optimization, which would suffer from the issue of stucking in the local minima. Therefore, the initialization, that takes into account the structure of the input graph, could play an important role in the learning of network embedding that largely enhances its performance.

\header {\bf Goal.} Given a graph $G=(V,E)$, we are to compute for each node $v \in V$ a coarse embedding $f(v)$, which preserves the sketching structure of $G$ and can be used as the initialization for the network embedding algorithms.

%% file: overview.tex
\section{Methodologies} \label{sec:overview}

A naive approach for the initialization of network embedding is by random, which assigns random numbers in ${\mathbb{R}}$ for the initial embedding of each node in the graph. However, this approach disregards the structure of the input graph, rendering it unsuitable for network embedding. Instead, we propose the graph partition based algorithm (GPA) that depicts the sketch of the input graph $G=(V,E)$ using the partitioning of $G$, which are then processed as the initial embedding of each node in $V$.

Specifically, GPA takes two phases in its computing framework, namely the preprocessing phase and the initialization phase, as shown in Figure~\ref{fig:framework}.

In the initialization phase, GPA first computes a partitioning $\mathcal{P}$ of $G$ by the graph partitioning algorithm, which produces $k$ disjoint subsets of $V$, where $k$ is a user-defined number and will be discussed in Section~\ref{sec:construct-abt}.
Then, we construct an abstract graph $G_a = (V_a, E_a)$ based on the partitioning of $G$, as aforementioned. Note that, the size of $G_a$ is $k$, which should be much smaller than the size of $G$, i.e., $|V_a| = k \ll |V|$.

After that, we compute the network embedding $f_a$ of the abstract graph $G_a$, which is a weighted graph, by a modified version of random walk based algorithm \cite{brs14}. Finally, each node in $G$ inherits the embedding of its corresponding abstract node in $G_a$, and then performs the embedding fusion among its neighbors via a propagation process. Once the propagation is converged, we obtain the initial embedding of each node, which will be taken as input by the network embedding algorithms on $G$.

On the other hand, in the preprocessing phase, we build a regression model that learns the configuration of hyperparameters for the network embedding algorithm on the abstract graph. As such, given an abstract graph, we are able to identify a suitable set of hyperparameters by inspecting the regression model with a linear time cost.

In what follows, we will elaborate the details of each step.

%% file: solution.tex
\subsection{Abstract Graph Construction} \label{sec:construct-abt}

To construct the abstract graph $G_a=(V_a, E_a)$ of $G=(V,E)$, we first obtain a partitioning $\mathcal{P}$ of $G$, denoted by $\mathcal{P} = \{V_1, V_2, \cdots, V_k\}$ where $k$ is a user-defined number. The goal of graph partition is $(k, \epsilon)$-balanced where $0 < \epsilon < 1$, such that it satisfies the constraint
\[
\max_{1 \leq i \leq k} |V_i| \leq (1+\epsilon) \Big\lceil \frac{|V|}{k} \Big\rceil,
\]
and also minimizes the size of edge-cut, i.e.,
\[
\bigcup_{1 \leq i, j \leq k} \{(v, u) \in E \mid v \in V_i, u \in V_j \}.
\]
However, the $(k, \epsilon)$-balanced graph partition is NP-hard \cite{ahip+16}. To address this issue, we resort to the METIS algorithm \cite{gv98} for graph partitioning, which is widely adopted in practice and incurs a running time complexity of $O(|V| + |E| + k \log k)$ \cite{gv05}.

Based on $\mathcal{P}$, we construct the abstract graph $G_a$ of $G$ by (i) creating an {\it abstract node} $v_a$ for each partition $V' \in \mathcal{P}$, i.e., $b(V') = v_a$, and (ii) connecting two abstract nodes $v_a$ and $u_a$ with an {\it abstract edge} $(v_a, u_a)$ of a weight $w(v_a,u_a)$ if and only if there exist $w(v_a,u_a) > 0$ edges $(v, u) \in E$ such that $v \in b^{-1}(v_a)$ and $u \in b^{-1}(u_a)$. Hence, the number of abstract nodes in $G_a$ is $k$, i.e., the number of partitions of $G$. Besides, the number of abstract edges of $G_a$ is bounded by the size of edge cut.

One crucial issue remaining is how to decide $k$. On one hand, if $k$ is small, then one abstract node would be pertinent to a lot of nodes in the input graph $G$. As such, the initial embedding of each node in $G$ inherited from the corresponding abstract node would lose the power of effectiveness. On the other hand, if $k$ is large, then the abstract graph $G_a$ would be large too. Therefore, it would be highly expensive to compute the network embedding on $G_a$, which increases the overall cost of the initialization phase. To strike a good balance, we set $k = \lceil \sqrt{|V|} \rceil$, which is a sufficiently large number but much smaller than $|V|$, that works well in practice.

\begin{algorithm}[t]
	\caption{Build-Alias($S$)}\label{alg:alias}
	\KwIn{The set $S$ of elements $e$ with the transition probability $P(e)$.}
	\KwOut{The alias probability $P_a(e)$ and the alias $A(e)$ for all $e \in S$.}

    Let $P_a(e) = |S| \cdot P(e)$ and $A(e) = e$\;
    Let $S_l = \{ e \in S | P_a(e) > 1 \}$ and $S_s = \{ e \in S | P_a(e) < 1 \}$\;
    \While{\textnormal{$S_l$ is not empty}}
    {
        Select any elements $x \in S_s$ and $y \in S_l$\;
        Let $A(x) = y$ and remove $x$ from $S_s$\;
        Decrease $P_a(y)$ by $1 - P_a(x)$\;
        \If{\textnormal{$P_a(y) \le 1$}}
        {
            Remove $y$ from $S_l$\;
            If $P_a(y) < 1$, then add $y$ into $S_s$\;
        }
    }
    \Return $P_a(e)$ and $A(e)$ for all $e \in S$.
\end{algorithm}

\subsection{Abstract Graph Embedding} \label{sec:abstract-emb}

To compute the network embedding $f_a$ of the abstract graph $G_a$, which is a weighted graph, we cannot directly exploit the previous network embedding techniques \cite{brs14,jmmm+15,aj16,jyhj+18} as they are tailored for the un-weighted graphs.

In order to remedy this issue, we adopt the random walk based algorithm, i.e., DeepWalk \cite{brs14}, with a slight modification to accommodate the network embedding learning on the abstract graph $G_a$. Note that, there are two phases of computation in the random walk based algorithms: First, it generates a number of random walks from each node in $G$; Then, it computes the embedding of each node by word2vec \cite{tikg+13}, which takes as input the random walks. There are some hyperparameters in the random walk based algorithms, namely the number of random walks and the length of a random walk, which would be configured by the hyperparameter learning module, as explained in the later section. While the second phase remains the same, the modification mainly happens in the first phase where the generation of random walks follows the distribution of weights on the abstract edges.

In particular, when generating the random walks on $G_a$, the transition probability of an edge $(u_a, v_a) \in E_a$, denoted by $P(u_a,v_a)$, is calculated as the fraction of the weight $w(u_a,v_a)$ among the total weights of the edges incident to $u_a$, i.e., $P(u_a,v_a) = \frac{w(u_a,v_a)}{\sum_{v_a' \in \nbr(u_a)} w(u_a, v_a')}$. Therefore, for each edge $(u_a,v_a) \in E_a$, we have (i) $0 < P(u_a,v_a) \leq 1$, and (ii) $\sum_{v_a' \in \nbr(u_a)} P(u_a, v_a') = 1$. In the generation of the random walk with the ending node $u_a$, we extend the walk by selecting a node $v_a \in \nbr(u_a)$ with the transition probability $P(u_a,v_a)$.

To make the selection of nodes in random walk efficiently, we resort to the alias method \cite{m91} with a preprocessing step, as illustrated in Algorithm~\ref{alg:alias}.
Specifically, the alias method builds for each element $e \in S$ an alias probability $P_a(e) \in [0, 1]$ and an alias $A(e) \in S$. To explain, for each element $e \in S$, the algorithm first enlarges the transition probability $P(e)$ by $|S|$ times, and sets the initial alias probability $P_a(e) = P(e) \cdot |S|$ and the initial alias of $e$ as itself (Line 1). Then, the algorithm works iteratively where each iteration selects two distinct elements $x$ and $y$ where $P_a(x) < 1$ and $P_a(y) > 1$, and then assigns $y$ as the alias of $x$ and decreases $P_a(y)$ by $1 - P_a(x)$. The algorithm terminates when there are no elements $y$ with $P_a(y) > 1$ (Lines 2-9). After that, to select an element from $S$, the alias method first randomly selects an element $e \in S$ with the probability $\frac{1}{|S|}$, and then chooses $e$ with the probability $P_a(e)$ or $A(e)$ with the probability $1-P_a(e)$. As a result, the time complexity of the preprocessing step and selecting an element is $O(|S|)$ and $O(1)$ respectively.

\begin{algorithm}[t]
	\caption{Propagate($G$, $f_a$, $\delta$)}\label{alg:propagate}

	\KwIn{The graph $G=(V,E)$, the embeddings $f_a$ of $G$'s abstract graph, and the threshold $\delta$.}
	\KwOut{The set $f_i$ of initial embedding of each node $v \in V$.}
    \SetKwRepeat{Do}{do}{while}

    Let $f_i(v) = f_a(p(v))$ for each node $v \in V$\;
    \Do {$\Delta > \delta$}
    {
        \For{{\bf each node } $v \in V$}
        {
            Let $f_{nbr}(v) = \frac{1}{|\nbr(v)|} \sum_{u \in \nbr(v)} f_i(u)$\;
            Compute $f_i'(v) = \frac{1}{2} ( f_i(v) + f_{nbr}(v) )$\;
        }
        Let $\Delta = \frac{1}{|V|} \sum_{v \in V} \| f_i'(v) - f_i(v) \|$\;
        For each node $v \in V$, let $f_i(v) = f_i'(v)$\;
    }
    \Return $f_i$.
\end{algorithm}

\begin{figure*}[t]
\centering
\begin{tabular}{c}
\includegraphics[width = 95mm]{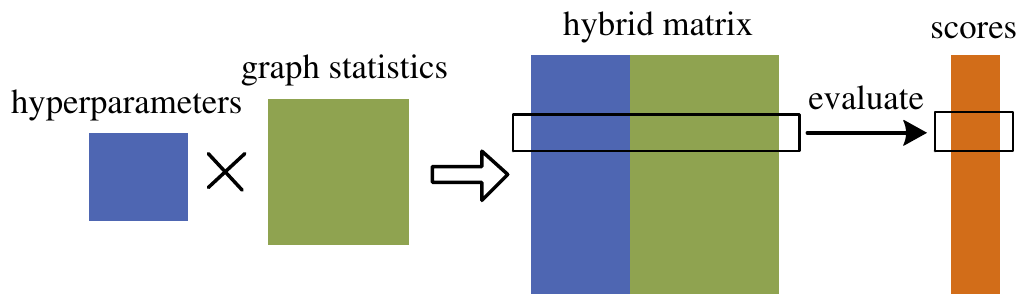}
\end{tabular}
\caption{The generation of training data for hyperparameter learning.} 
\label{fig:hyperparameter}
\end{figure*}

\subsection{Embedding Propagation} \label{sec:progation}

To compute the initial network embedding of $G$ from the network embedding $f_a$ of the abstract graph $G_a$, a naive approach is to let the initial embedding of each node $v$ equal the embedding of the corresponding abstract node $p(v)$. However, this approach would suffer from the issue where the nodes pertinent to the same abstract node have the same initial embeddings, rendering this approach ineffective.

In order to address this issue, we devise an iterative approach where each node updates its own embedding based on the embeddings of its neighbors until the convergence is met. Specifically, in each iteration, each node $v \in V$ first aggregates the embeddings of $v$'s neighbors, which results in the average embedding $f_{nbr}(v)$. Then, we update $v$'s embedding as the aggregation of $f_{nbr}$ and its own embedding $f_i(v)$. The rationale is that the embedding of a node should be close to the ones of its neighbors in the graph.

Algorithm~\ref{alg:propagate} illustrates the procedure of embedding propagation. Consider a graph $G = (V,E)$, the abstract graph $G_a$ of $G$, and the network embedding $f_a$ of $G_a$. At the beginning, for each node $v \in V$, we let the initial embedding $f_i(v)$ of $v$ be the embedding $f_a(p(v))$ of its abstract node $p(v)$ in $G_a$ (Line 1). Then, the algorithm works in several iterations. In each iteration, the updating of the embedding of each node $v \in V$ can be achieved in a two-layer computing framework. In the first layer, we compute the average embedding $f_{nbr}$ among its neighbors (Line 4), i.e.,
\[
f_{nbr}(v) = \frac{1}{|\nbr(v)|} \sum_{u \in \nbr(v)} f_i(u).
\]
Then, we employ another layer to calculate the updated embedding $f_i'(v)$ of $v$ as the average of $f_i(v)$ and $f_{nbr}(v)$, i.e.,
\[
f_i'(v) = \frac{1}{2} ( f_i(v) + f_{nbr}(v) ).
\]
After that, for all nodes $v \in V$, we compute the average difference between the updated embedding $f_i'(v)$ and the previous embedding $f_i(v)$ on their Euclidean distance (Line 6), denoted by
\[
\Delta = \frac{1}{|V|} \sum_{v \in V} {\| f_i'(v) - f_i(v) \|}.
\]
Now, we can update the embedding $f_i(v)$ of $v$ as $f_i'(v)$, i.e., $f_i(v) = f_i'(v)$, which completes this iteration. If the average difference $\Delta$ is not more than a user-defined threshold $\delta$, we terminate this procedure and return $f_i$ as the result. Otherwise, we continue updating the embedding of each node $v \in V$ until convergence is met, i.e., $\Delta \leq \delta$. Note that, $\delta$ is usually set as a value proportional to $\frac{1}{|V|}$. Consequently, the time complexity of one iteration is $O(\sum_{v \in V} |\nbr(v)|) = O(|E|)$, as each node needs to inspect the embeddings of its neighbors once.

\begin{table}[t]
\centering
\caption{Hybrid features for hyperparameter learning.}\label{tbl:hyper-feature}
\begin{tabular}{l|l} \hline
Category & Feature \\ \hline \hline
\multirow{2}{*}{hyperparameters} & the number of random walks \\ \cline{2-2}
 & the length of a random walk \\ \hline
\multirow{8}{*}{graph statistics} & the number of nodes of $G_a$ \\ \cline{2-2}
 & the number of edges of $G_a$ \\ \cline{2-2}
 & the density of $G_a$ \\ \cline{2-2}
 & the diameter of $G_a$ \\ \cline{2-2}
 & the average degree of $G_a$ \\ \cline{2-2}
 & the maximum degree of $G_a$ \\ \cline{2-2}
 & the average edge weight of $G_a$ \\ \cline{2-2}
 & the maximum edge weight of $G_a$ \\ \hline
\end{tabular}
\end{table}

\subsection{Hyperparameter Learning} \label{sec:hyperparameter}

There is one crucial issue remaining in the network embedding learning on the abstract graph $G_a$ which is the configuration of hyperparameters in the random walk based algorithm, i.e., the number of random walks and the length of a random walk. A naive approach is to configure the hyperparameters with random values. However, this approach would severely degrade the performance of the network embedding algorithm. Alternatively, one might propose the solution that exploits the existing optimization techniques \cite{jryb11}, such as grid search, to tune the hyperparameters on the fly. Nevertheless, this approach would greatly increase the running time of the network embedding algorithm, as the optimization could be costly.

To cope with this issue, we utilize a preprocessing phase which trains a regression model that takes into account both the hyperparameters and the statistics of the abstracts graphs. As such, given an abstract graph $G_a$, we are able to infer from the model the suitable hyperparameters for $G_a$ with a slight cost, as explained shortly.

Table~\ref{tbl:hyper-feature} shows the hybrid features for hyperparameter learning, which consists of two features from the category of hyperparameters and eight features from the category of graph statistics.

As illustrated in Figure~\ref{fig:hyperparameter}, to generate the training data with the hybrid features, we first construct a set $\mathcal{H}$ of hyperparameter combinations and a set $\mathcal{S}$ of graph statistics for each abstract graph $G_a$. Specifically, we enumerate the possible values for each hyperparameter by heuristic to produce the set $\mathcal{H}$. Besides, to generate $\mathcal{S}$, we first exploit the random graph generation technique \cite{jdjc05} to generate a set $\mathcal{G}$ of random abstract graphs. And then, we utilize the graph mining tool, SNAP \cite{ls16}, to calculate the statistics of each graph in $\mathcal{G}$, which results in the set $\mathcal{S}$. After that, for each hyperparameter combination $H \in \mathcal{H}$ and each graph statistics $S \in \mathcal{S}$, we concatenate $H$ and $S$ to generate one data point with the hybrid features. That is, the total number of data points will be $|\mathcal{H}| \cdot |\mathcal{G}|$. All data points with the hybrid features together form the hybrid matrix, denoted by $\boldsymbol{X}$.

For each row in the hybrid matrix $\boldsymbol{X}$, which is generated from a hyperparameter combination $H$ and the statistics $S$ of an abstract graph $G_a$, we compute the network embedding $f_a$ on $G_a$ with hyperparameters in $H$. Then, we evaluate $f_a$ on Equation~\ref{eqn:goal} with a slight modification where $\theta$ is an Euclidean distance function and $\boldsymbol{A}_{v,u}$ is $w(v,u)$ if $(v, u) \in E_a$ otherwise $0$. As such, for all the data points in $\boldsymbol{X}$, we obtain a vector of the evaluation scores, denoted by $\boldsymbol{Y}$.

Hence, our goal is to find a vector $\boldsymbol{w}$, such that we have
\[
\boldsymbol{X} \cdot \boldsymbol{w}^T = \boldsymbol{Y}.
\]
As a result, the objective is
\[
\min_{w_1, w_2, \dots, w_{\alpha}} \sum_{1 \leq i \leq \beta} {( \sum_{1 \leq j \leq \alpha}{x_{ij} \cdot w_j} - y_i )^2}
\]
where $\alpha$ is the number of dimensions of $\boldsymbol{X}$ and $\beta$ is the number of data points in $\boldsymbol{X}$. Solving the above formula by stochastic gradient descent, we are able to identify the vector $\boldsymbol{w}$ that largely approximates to the optimal solution.

Once obtained the regression model, i.e., $\boldsymbol{w}$, we can compute a suitable configuration of hyperparameters for a given abstract graph $G_a$ efficiently. To explain, we first produce the graph statistics $S$ of $G_a$ by utilizing SNAP. Then, we inspect each hyperparameter combination $H \in \mathcal{H}$, and generate a data point $\boldsymbol{x}$ with the hybrid features by concatenating $H$ and $S$. Hence, we can calculate the score of the data point $\boldsymbol{x}$ as $y = \boldsymbol{x} \cdot \boldsymbol{w}^T$. In the end, we choose the hyperparameter combination $H \in \mathcal{H}$ with the highest score. Note that, the time complexity of identifying the suitable hyperparameters is $O(|\mathcal{H}|)$.

%% file: experiments.tex
\begin{table}[t]
\centering
\caption{Datasets.}\label{tbl:exp-data}
\begin{tabular}{l|c|r|r|r} \hline
Dataset & Category & \#Nodes & \#Edges & \#Labels \\ \hline \hline
{\it Enron}\footnotemark[1]  & email         & 36,692    & 183,831   & 0 \\ \hline
{\it GRQC}\footnotemark[2]    & collaboration & 5,242     & 14,496    & 0 \\ \hline
{\it Blog}\footnotemark[3] & social    & 10,312    & 333,983   & 39 \\ \hline
{\it Wiki}\footnotemark[4] & word        & 4,777     & 184,812   & 40 \\ \hline
\end{tabular}
\end{table}

\footnotetext[1]{\scriptsize http://www.cs.cmu.edu/$\sim$enron}
\footnotetext[2]{\scriptsize http://snap.stanford.edu/data/ca-GrQc.html}
\footnotetext[3]{\scriptsize http://socialcomputing.asu.edu/datasets/BlogCatalog}
\footnotetext[4]{\scriptsize www.mattmahoney.net/dc/textdata}

\section{Experimental Evaluations} \label{sec:experiments}

\begin{table*}[t]
\centering
\caption{Precisions in the task of link prediction evaluated by Cosine similarity and Euclidean similarity.}\label{tbl:exp-link-prediction}
\begin{tabular}{c|c|c|c|c|c|c|c|c|c} \hline
\multirow{2}{*}{Algorithm} & \multirow{2}{*}{Initialization} & \multicolumn{4}{c|}{Cosine Similarity} & \multicolumn{4}{c}{Euclidean Similarity} \\ \cline{3-10}
 & & Enron & GRQC & Blog & Wiki & Enron & GRQC & Blog & Wiki  \\ \hline

\multirow{3}{*}{node2vec} 
& GPA & {\bf 0.9579}   &	{\bf 0.9933}  &	{\bf 0.9816} & {\bf 0.9325}   &   {\bf 0.9665}	&  {\bf 0.9947} &	{\bf 0.9887} & {\bf 0.9438} \\ \cline{2-10}
& HARP  & 0.9209	& 0.9621	&    0.9708 & 0.9210 & 0.9418	&   0.9846	&   0.9618 & 0.9258 \\ \cline{2-10}
& Random & 0.9136	& 0.9533	&    0.9631 & 0.9117 & 0.9309	&   0.9817	&   0.9587  & 0.9217 \\
\hline

\multirow{3}{*}{DeepWalk} 
& GPA & {\bf 0.9702}    &	{\bf 0.9937} &	{\bf 0.9820} & {\bf 0.9315}   & {\bf 0.9691}    &	{\bf 0.9958}    &	{\bf 0.9879} & {\bf 0.9411} \\ \cline{2-10}
& HARP  & 0.9352 &	0.9625    &	0.9717 & 0.9178    &   0.9449  &	0.9842    &	0.9658 & 0.9354 \\ \cline{2-10}
& Random & 0.9218 &	0.9430    &	0.9535  & 0.9024  &   0.9355  &	0.9764    &	0.9517 & 0.9276 \\
\hline

\multirow{3}{*}{LINE} 
& GPA & {\bf 0.7849}    &	{\bf 0.9852}    &	{\bf 0.9436}  & {\bf 0.8175}   &  {\bf 0.5790}   &	{\bf 0.9665}    &	{\bf 0.9274} & {\bf 0.8356} \\ \cline{2-10}
& HARP  & 0.7484 &	0.9526    &	0.9298  & 0.7849  &   0.5372  &	0.9471    &	0.9016  & 0.8126 \\ \cline{2-10}
& Random & 0.7414 &	0.9411    &	0.9127   & 0.7658  &   0.5237  &	0.9392    &	0.8836  & 0.8028 \\
\hline

\end{tabular}
\end{table*}

In this section, we demonstrate that the proposed graph partition based algorithm, dubbed as GPA, outperforms the state-of-the-art, i.e., HARP \cite{hbys18}, as well as the randomized method, denoted by Random, on various datasets and over different tasks, such as link prediction and node classification. In particular, we apply the initialization techniques of GPA, HARP and Random to the widely-used network embedding algorithms, i.e., node2vec \cite{aj16}, DeepWalk \cite{brs14}, and LINE \cite{jmmm+15}.
Note that, (i) the original versions of network embedding algorithms adopt Random as its initialization method, and (ii) for each algorithm, we set the embedding vector size $d=128$ and their other hyperparameters as the recommended ones in all experiments.


Our algorithms are implemented in Scala and C++, and all experiments are conducted on a machine with 8 GB memory and an Intel Core i5 CPU (2.3 GHz), which is installed with the macOS. For each set of experiments, we perform each algorithm 10 times and report the average reading.


Following the previous work \cite{aj16,ls16}, we evaluate the performance of the proposed algorithms against $4$ datasets from various categories in our experiments, as shown in Table~\ref{tbl:exp-data}.

\subsection{Evaluations on Link Prediction}

In the first set of experiments, we evaluate the performance of network embedding with the initialization, provided by GPA, on the task of link prediction. Specifically, we compare GPA against HARP and Random on the graphs: Enron, GRQC, Blog, and Wiki.

To generate the testing and training sets for the task of link prediction on each graph $G=(V,E)$, we first randomly select $\lceil \alpha |E| \rceil$ number of edges from $E$, denoted by $E_s$, where $0 < \alpha < 1$. Then, we remove $E_s$ from $E$, resulting in the residual set $E_r$ of edges, i.e., $E_r = E \setminus E_s$. After that, we compute the largest connected component $C$ of the graph induced on the edges in $E_r$. Finally, we produce the training set consisting of the edges in $E_r$ whose nodes are in $C$, and generate the testing set that contains two parts: (i) The positive samples, i.e., the set of the edges of $E_s$ whose nodes are both in $C$, and (ii) the negative samples, i.e., the set of random pairs of nodes $u$ and $v$ in $C$ where $(u,v)$ is not an edge in $E$. Note that, in the experiments, we set $\alpha = 10\%$ and the size of testing set as $2|E_s|$, i.e., the number of positive samples equals the number of negative samples. 
Additionally, due to practical considerations, for each node $v$ appearing in $E_s$, the number of positive samples incident to $v$ should be equal to the number of negative samples incident to $v$.

For each graph $G=(V,E)$, we compute the embedding of each node in $V$ by running the network embedding with the initialization techniques on the training set, and then calculate the similarity of all pairs of nodes in the testing set. For each node $v$, we predict the top $t$ nodes that are the most similar to $v$, where $t$ is the number of positive samples incident to $v$ in $E_s$. 
We adopt two kinds of similarity measures: Cosine similarity and Euclidean similarity. Given two vectors $\boldsymbol{x}$ and $\boldsymbol{y}$ of the same length, the Cosine similarity of $\boldsymbol{x}$ and $\boldsymbol{y}$ is $\frac{ \boldsymbol{x} \cdot \boldsymbol{y}}{ \| \boldsymbol{x} \| \| \boldsymbol{y} \| }$, and the Euclidean similarity of them is $\| \boldsymbol{x} - \boldsymbol{y} \|$. In the end, we calculate the accuracy as the fraction of positive samples in the most similar $|E_s|$ pairs of nodes in the testing set.

Table~\ref{tbl:exp-link-prediction} shows the accuracy of node2vec, DeepWalk, and LINE with the initialization techniques, i.e., GPA, HARP, and Random, for link prediction by Cosine similarity and Euclidean similarity on the datasets Enron, GRQC, Blog, and Wiki respectively. As we can see, GPA outperforms HARP on all datasets and in terms of both similarity measures, and the results of HARP is slightly better than the ones of Random. In particular, on the Enron dataset using the Euclidean similarity, GPA is better than HARP on LINE by $7.8\%$, on node2vec by $2.6\%$, and on DeepWalk by $2.5\%$.
This is due to that GPA exploits several effective strategies that overcome the shortage of HARP and lead to a better initialization for the network embedding algorithms.

\begin{table*}[t]
\centering
\caption{F1 scores in the task of node classification.}\label{tbl:exp-node-classification}
\begin{tabular}{c|c|c|c|c|c} \hline
\multirow{2}{*}{Algorithm} & \multirow{2}{*}{Initialization} & \multicolumn{2}{c|}{Micro-F1 score} & \multicolumn{2}{c}{Macro-F1 score} \\ \cline{3-6}
 & & Blog & Wiki & Blog & Wiki  \\ \hline

\multirow{3}{*}{node2vec} & GPA & {\bf 0.3174}    &	{\bf 0.6310}    & {\bf 0.2395}    &	{\bf 0.5830} \\ \cline{2-6}
& HARP & 0.3028	&   0.6192  &  0.2281   &	0.5631 \\ \cline{2-6}
& Random & 0.2916	&   0.6033  &  0.2195   &	0.5587 \\
\hline

\multirow{3}{*}{DeepWalk} & GPA & {\bf 0.3399}    &	{\bf 0.6295}    & {\bf 0.2563}    &	{\bf 0.5616} \\ \cline{2-6}
& HARP & 0.3191 &	0.6029    & 0.2387    &	0.5481 \\ \cline{2-6}
& Random & 0.3106 &	0.5967    & 0.2315    &	0.5380 \\ 
\hline

\multirow{3}{*}{LINE} & GPA & {\bf 0.3070}    &	{\bf 0.4987}    & {\bf 0.2082}    &	{\bf 0.4282} \\ \cline{2-6}
& HARP & 0.2823 &	0.4798    &  0.2029   &	0.4165 \\ \cline{2-6}
& Random & 0.2799 &	0.4687    &  0.1982   &	0.4091 \\\hline

\end{tabular}
\end{table*}

\subsection{Evaluations on Node Classification}

In node classification, we evaluate the performance of GPA, HARP and Random on the datasets, i.e., Blog and Wiki, whose nodes are associated with labels. 
We run the embedding algorithm with the initialization techniques on each graph to obtain the embedding of nodes, which are then input to a multi-class logistic regression classifier utilizing one-vs-rest technique and L2 regularization. We randomly split the set of nodes equally to generate the training and testing sets respectively. Following the previous work \cite{aj16}, we measure the performance of GPA, HARP, and Random in micro-F1 score and macro-F1 score.

Table~\ref{tbl:exp-node-classification} presents the micro-F1 score and macro-F1 score of all the algorithms on the datasets Blog and Wiki. Observe that GPA consistently outperforms HARP in all settings, and HARP is slightly better than Random. In particular, regarding the method of LINE, the relative performance gain on Blog of GPA compared to HARP is $8.76\%$ in micro-F1 score and $2.62\%$ in macro-F1 score. Besides, on Wiki, DeepWalk with GPA gives us $4.41\%$ gain in micro-F1 score and $2.46\%$ gain in macro-F1 score. This again demonstrates the superiority of our graph partition based approach that provides effective initialization for network embedding.

\subsection{Evaluations on Efficiency}

\begin{figure*}[t]
\centering
\begin{tabular}{c}
\includegraphics[height=45mm]{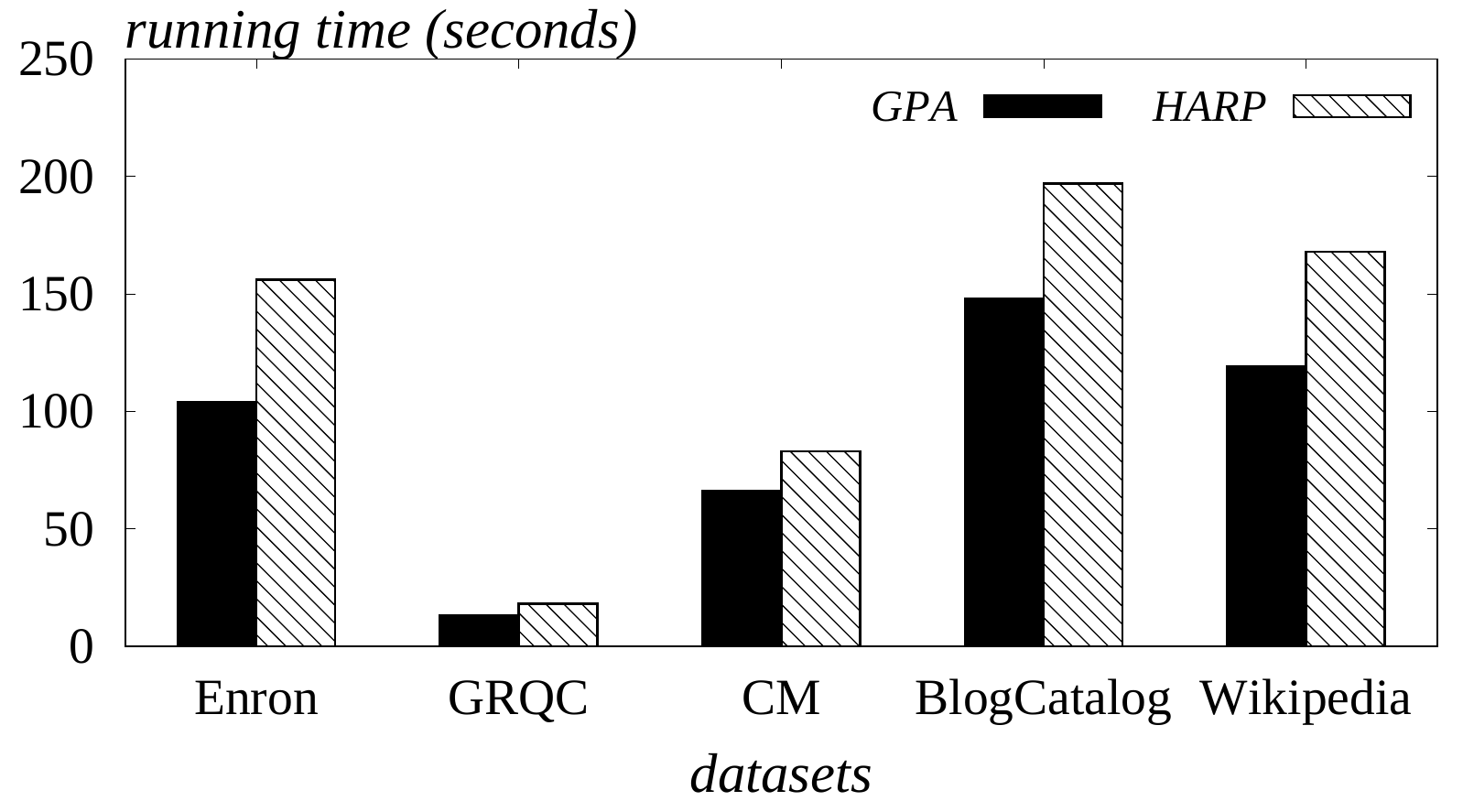}  \\ 
\end{tabular}
\caption{The running time of the initialization techniques.}
\label{fig:exp-time}
\end{figure*}

In this experiment, we evaluate the efficiency of GPA by comparing with HARP on all datasets. Figure~\ref{fig:exp-time} reports the running time of GPA and HARP that take as input the whole graph in each dataset. GPA is much faster than HARP on all datasets with at least 20\% performance gain. In particular, GPA reduces the running time by 33.33\% compared to HARP on the Enron dataset. This is because HARP computes the initial embedding of each node in a hierarchical manner that requires several iterations of computation, while GPA reduces the input graph to the abstract graph of size $\large \lceil \sqrt{n} \large \rceil$ whose embeddings are then propagated among the nodes in the input graph with a linear cost, where $n$ is the number of nodes in the input graph.

%% file: related.tex
\section{Related Work} \label{sec:related}

Network embedding or graph representation learning has been intensively studied in the literature (see \cite{hvk17,pxjw17,wrj17,djxc18,pe18,hbrs18} and the references therein). Most of these approaches \cite{brs14,jmmm+15,aj16,wzj17,jyhj+18} exploit negative sampling or skip-gram models, which turn out to be the non-convex problem \cite{yo14,hbys18} and usually solved by stochastic gradient descent (SGD). However, few of them takes into account the effect of the initial embedding of each node in the network that would dramatically impact the performance of the algorithms.

Besides HARP \cite{hbys18}, introduced in Section~\ref{sec:introduction}, MILE \cite{jss18} also adopts the hierarchical computing framework, almost the same as HARP, but differs from HARP in that it aims to compute the final network embedding for the input graph.

On the other hand, Mishkin et al. \cite{dj15} discussed the importance of initialization in the training of deep neural networks. However, their approach does not consider the graph data, and can not be applied to network embedding. The other line of research on network embedding is for different problem setting or datasets \cite{czdm+17,kpxf+18,yzzj+18}, making them unsuitable for solving the problem of this paper.

%% file: conclusions.tex
\section{Conclusions} \label{sec:conclusions}

In this paper, we studied the issue of initialization for network embedding that would significantly affect the performance of network embedding algorithms. To address this issue, we proposed the algorithm GPA that constructs the abstract graph sketching the input graph by well partitioning the input graph. We developed a weighted network embedding algorithm to compute the embedding of nodes in the abstract graph. After that, the network embedding of the abstract graph will be propagated among the nodes of the input graph, which leads to the initial embedding of the input graph. Besides, to make the weighted network embedding algorithm efficient, we devised a regression model to address the issue of hyperparameter tuning in the weighted network embedding algorithm. Finally, we demonstrated the effectiveness and efficiency of GPA against the state-of-the-arts on various datasets. In particular, GPA achieves the performance gains of up to 7.76\% and 8.74\% on link prediction and node classification respectively, and reduces the running time by at least 20\%.